\documentclass[aps,reprint,amsmath,amssymb]{revtex4-2}
\usepackage{graphics}
\usepackage{subfigure}
\usepackage{epsfig}
\usepackage{epsf,epic}
\usepackage{color}
\usepackage{marvosym}
\usepackage{subfigure}
\usepackage{amsmath}
\usepackage{amssymb}
\usepackage{amsfonts} 
\usepackage{wrapfig}
\usepackage{pstricks}
\usepackage{multirow}
\usepackage{bm}
\graphicspath{{./figs/}}
\usepackage{dcolumn}
\newcommand{\etal}{\textit{et al.\ }}

\begin{document}
\title{Calculated electron paramagnetic resonance $g$-tensor and hyperfine
  parameters for zinc vacancy and N related defects in ZnO}

\author{Klichchupong Dabsamut$^{a,b}$}\email{klichchupong.d@gmail.com}
\author{Adisak Boonchun$^{b}$}
\author{Walter R. L. Lambrecht $^a$}\email{walter.lambrecht@case.edu}
\affiliation{$^a$ Department of Physics, Case Western Reserve University, 10900 Euclid Avenue, Cleveland, Ohio 44106-7079, USA}
\affiliation{$^b$ Department of Physics, Faculty of Science, Kasetsart University, Bangkok 10900, Thailand}

\begin{abstract}
  Various defects in ZnO, focused on substitutional N$_\mathrm{O}$ and N$_2$ in various sites, O-site, interstitial and Zn-site are studied using first-principles calculations with the goal of understanding the electron
  paramagnetic resonance (EPR) center reported for N$_2$ in ZnO and substitutional N on the O-site.
  The $g$ tensors are calculated using the gauge including
  projector augmented wave (GIPAW) method and compared with experiments.  The $g$-tensor of the
  free N$_2^+$ and N$_2^-$ radicals and their various contributions within the GIPAW theory 
  are analyzed first to provide a baseline reference for the accuracy of the method and for understanding the
  N$_2$ behavior in ZnO. 
  Previous controversies on the site location of N$_2$ in ZnO for this EPR center and on the shallow or deep nature and donor or acceptor nature of this center are resolved. We find that the N$_2$ on the Zn site is mostly zinc-vacancy like in its spin density and $g$-tensor, while for the O-site, a model with the N$_2$ axis lying  in-the basal plane  and the singly occupied $\pi_g$-orbital along the {\bf c} axis
  provides good agreement with experiment. For the interstitial location, if the N$_2$ is not strongly interacting with the surroundings, no levels in the gap are found and hence also no possible EPR center.
  The calculated $g$-tensors for N$_\mathrm{O}$
  and $V_\mathrm{Zn}$ are also found to be in good agreement with experiment. The effects of different functionals
  affecting the localization of the spin density are shown to affect the $g$-tensor values.
\end{abstract}

\maketitle
\section{Introduction}
Electron Paramagnetic Resonance (EPR) provides  one of the most
powerful methods  to study defect electronic structure.
The $g$-tensor which describes the spin-splitting
of a defect level in a magnetic field as function of the magnetic field
magnitude and direction provides a unique fingerprint for the defect. Along with the
hyperfine tensor, which describes the interaction with nuclear spins associated with the defect, the chemical identity of a defect can readily be determined.
In combination with optical or thermal excitation or quenching of the
EPR center, information on the defect levels 
can be obtained. However, $g$-tensors are rarely calculated
from first-principles.  Defects are usually described in periodic boundary
conditions and it is only fairly recently that the methodologies
for calculating $g$-tensors were developed for periodic systems.
It requires calculating the induced current
response to an external magnetic field or the orbital magnetization.  This nontrivial problem was
first solved for nuclear magnetic resonance (NMR) chemical shielding factors
by including the gauge induced changes in the phases of the wave function
in work by Mauri \etal\cite{Mauri96,Mauri97}.
Subsequently,  Ziegler  and coworkers developed these approaches
in the context of atom centered basis sets
 \cite{Skachkov2010,Kadantsev2009} and 
 Mauri \etal developed an implementation
 in terms of the projector augmented wave methods,\cite{Pickard01,Pickard02} known as the GIPAW (gauge including projector augmented wave) method.
 More recently,
 Ceresoli developed a non-perturbative approach based on Berry phases\cite{Ceresoli10} and also further improved the {\sc GIPAW} code. The GIPAW method was applied
 to a number of defect systems by Gerstmann \etal\cite{Gerstmann07,Gerstmann10,Pfanner12} and Skachkov \etal \cite{Skachkov2019ga2o3,Skachkov2019mg,Skachkov2020}.
 These works illustrate the capability of the combination of theory and
 experiment in EPR to distinguish various defect models for a
 given EPR center.

 Here we apply the GIPAW method to the study of several defects in ZnO.
 Our initial motivation was the work   by
 Garces \etal\cite{Garces03} identifying N$_2$ in ZnO.
 They identified N$_2$ unequivocally on the basis of
 the characteristic hyperfine interaction with two $I=1$  N nuclei and
 hypothesized that the N$_2$ occurred on an O-site. They found an axially
 symmetric $g$-tensor with the symmetry axis along the $c$-axis of the wurtzite structure of ZnO.  Subsequently,
 a computational study by Boonchun and Lambrecht\cite{Boonchun13}
 proposed instead a  Zn-vacancy location for the N$_2$ based 
 on the fact that the $g$-tensor agrees more closely with that
 of a N$_2^+$ radical than that of a N$_2^-$ radical.  In fact,
 when N$_2$ sits on a Zn-site, it behaves as double acceptor with
 the 10 valence electrons of N$_2$ compared to the 12 valence electrons
 of Zn (including the filled $3d$ shell). The N$_2$ molecule, which then plays the role  of a 2+ ion, would miss two electrons from its HOMO (highest occupied molecular orbital) $\sigma_{g+}$ level.  The $q=-1$ state of the defect
 then corresponds
 to the singly occupied $\sigma_{g+}$ state ( or a N$_2^+$ radical)
 and is EPR active. The $g$-tensor
 of this N$_2^+$ radical was calculated by Bruna and Grein\cite{Bruna08}
 and is characterized by a negligible $\Delta g_\parallel$-shift from the free electron value in the direction parallel to the bond and negative $\Delta g_\perp$
 in the direction
 perpendicular to the bond. This is readily understood in terms of
 second-order perturbation theory in which the $\Delta g$-tensor  arises
 from the cross effect of spin-orbit coupling and the orbital Zeeman effect
 and can be written as
 \begin{equation}
   \Delta g_{ij}= 2\lambda \sum_n \frac{\langle0|L_i|n\rangle\langle n|L_j|0\rangle}{E_0-E_n}, \label{eqgper}
 \end{equation}
 where $\lambda$ is the atomic spin-orbit coupling, $|0\rangle$
 is the singly occupied molecular orbital (SOMO) whose spin splitting we try
 to calculate and $|n\rangle$ are the other states with
 energy $E_0$ and $E_n$ respectively and, $L_i$ and $L_j$ are the cartesian components
 of the angular momentum operator.
 Since the angular momentum matrix elements from the SOMO  $\sigma_{g+}$
 state can here only couple to the higher lying $\pi_g$ LUMO (lowest unoccupied  molecular orbital) for the components perpendicular to the axis
 of the molecule, this gives a negative contribution to the $\Delta g_\perp$
 as was indeed observed in the work of Garces \etal\cite{Garces03}.
 Boonchun and Lambrecht estimated this $\Delta g_\perp$ for the N$_2$ molecule
 using a tight-binding model for the N$_2$ molecule with parameters fitted
 to density functional theory (DFT) calculation and using a
 calculated atomic spin-orbit coupling
 parameter to be  $-2600$ ppm in excellent agreement with Bruna and Grein's \cite{Bruna08}
 calculation which gave a value of  $-2734$ ppm. Both are in good agreement
 with the angular average of $\Delta g$ which amounts to $\sim (2/3)\Delta g_\perp$ and experimentally is about $-1900$ ppm.  
 The latter calculation
 was based on a more advanced quantum chemical calculation of the molecular
 levels but used a similar perturbation theoretical approach. 

 On the other hand, on an O-site one would expect the N$_2$ molecule
 to behave as a donor with an additional electron in the $\pi_g$
 state of the molecule, which then becomes a N$_2^-$ radical.
 One then would expect a positive
 $\Delta g_\perp$-tensor within the same type of perturbation theory,
 as we'll show explicitly later in Sec. \ref{sec:radical}.
 The $g$-tensors of N$_2^-$ on anion sites
 in MgO and KCl and other ionic compounds are well-known \cite{Napoli10}. They 
 are a bit more complex because the crystal environment breaks
 the degeneracy  of the $\pi_g$ state \cite{Napoli10}. The main argument
 of Boonchun and Lambrecht \cite{Boonchun13}
 was that the $g$-tensors of N$_2$ occurring
 on anion sites in these crystals differs significantly from that
 observed by Garces \etal\cite{Garces03}. However, in retrospect, it seems
 somewhat inconsistent that to explain the size of the hyperfine
 splitting one needs to  assume a significant delocalization of
 the spin density beyond the molecule while  for the $g$-tensor these models
 focused exclusively on the isolated molecule. Also, besides spin-orbit and
 orbital Zeeman perturbations, the full theory of
 $g$-tensors as implemented in the {\sc GIPAW} code
 includes additional contributions,
 such as the spin-other-orbit terms which involve the magnetic field induced
 by the first-order induced current, diamagnetic contributions and so on. 
 It  seems worthwhile applying this method to re-evaluate the $g$-tensor
 for N$_2$ in ZnO.
 
 The proposal by Boonchun and  Lambrecht \cite{Boonchun13} that N$_2$ on Zn would be a relatively shallow acceptor was
 exciting  because this could potentially lead to a path to
 the $p$-type doping of ZnO, which remains a challenging problem till today \cite{Reynolds13,Reynolds14}.
 However, their proposal was challenged in several ways. Petretto and Bruneval
 \cite{Petretto14}  found that the N$_2$ in the neutral state prefers
 to make a bridge type bond to two of the surrounding O atoms while
 the $q=-1$ state of this molecule prefers the isolated site similar to
 the calculation of Boonchun and Lambrecht. However, this much lower
 energy of the $N_2$ neutral state than leads to a much deeper $0/-$
 transition level making the system a deep rather than shallow acceptor.
 Furthermore they showed that energetically N$_2$ prefers the O site over the
 Zn-site.   Earlier, Nickel and Gluba \cite{Nickel09} found several 
 N$_2$ interstitial sites in ZnO to have lower energy than on the O-site. 

 The claim of a shallow acceptor behavior  of N$_2$ in ZnO was also challenged
 by an experimental study by Phillips \etal \cite{Phillips14} which studied the recharging behavior of the  EPR active state. This study like Garces \etal\cite{Garces03} found that upon irradiation with light, above a critical phonon energy of about 1.9 eV  a new signal identified with N$_\mathrm{O}$ becomes
 activated but
 unlike the Garces \etal  study it also found the N$_2$ signal to increase
 already at 1.4 eV while Garces \etal found irradiation to quench the EPR signal
 of N$_2$. To explain this, they proposed that their  sample could be
 inhomogeneous with
 different Fermi level positions in different regions of the sample placing
 the Fermi level close to the defect level of the N$_2$, whereby not all
 N$_2$ centers would originally be in the EPR active state. They associate the 1.4 eV activation energy with a transition from the defect to the conduction band
 and thus concluded the levels were deep. While they do not 
 explicitly discuss which site the N$_2$ is located on, this also suggests that the EPR active state is in a positive charge state because it requires removing an electron from the defect to the conduction band, and 
 hence that N$_2$ in ZnO is donor like.  That would, in fact, correspond to Garces \etal \cite{Garces03}'s proposal. However, alternative explanations for the
 recharging behavior could still be possible and  an explanation for the
 $g$-tensor itself is lacking.

 From the above it is clear that several open questions remain
 on N$_2$ in ZnO. This makes it worthwhile to revisit the
 N$_2$ calculations in different sites, O, interstitial and Zn and
 explore whether different orientations of the molecule can occur.
 Calculating the $g$-factors should help to identify which of these
 various possible models corresponds to the experimental EPR signal
 and the corresponding energy levels at the hybrid functional level
 can be compared with experiment. 
 To complement this study we also calculate the N$_\mathrm{O}$ $g$-tensor.
 We start from first-principles calculations for both the N$_2^-$ and N$_2^+$
 molecules and compare these with previous calculations as a test of the accuracy of the method. 

 As we will show, only the N$_2$ on the O site originally proposed
 by Garces \etal \cite{Garces03}
 has a clear spin localization on the N$_2$ molecule.
 The other systems have spin densities mostly on surrounding O atoms
 or very delocalized spin density.  This suggests that the N$_2$ on Zn-site
 electronic structure is closely related to the $V_\mathrm{Zn}$.
 We thus also calculate the $g$-factor
 for the Zn-vacancy, for which also experimental data are available.
 Good agreement with these experimental $g$-tensor data is established. 
 For the interstitial location, in models in which the N$_2$ minimally perturb the system, we find no levels in the gap and hence the $+$ charge state corresponds  to removing charge from the valence band maximum, leading to a
 very delocalized spin density and $g$ tensors not compatible with the
 experimental data for N$_2$ in ZnO.

 We will show that the N$_2$ on O site can explain
 qualitatively the data if we assume that the experiment measures
 some unresolved average over different symmetry equivalent orientations of the defect.  The results are also sensitive to the density functional
 used as discussed in the computational methods section.  We also calculate
 the $g$-factor for the substitutional N on O case and find reasonable agreement
 assuming again some degree of averaging occurs in the experiment.
 These results indicate that it might be possible to further resolve these
 EPR signals into separate centers corresponding to different orientations of the electronic structure on symmetry equivalent orbitals in future work, perhaps using higher magnetic fields and microwave frequencies to improve the resolution.

 \section{Computational method}\label{sec:comp}
 The initial defect
 relaxations are carried out at the hybrid functional level using a
 parametrized Heyd-Scuseria-Ernzerhof (HSE) potential \cite{HSE03,HSE06}
 in which the
 fraction of exact exchange $\alpha$ is set to 0.375 and the standard screening
 length $\mu=10$ \AA\ is used to cut off the long-range part of exact exchange.
 The relaxation calculations were carried out using the {\sc VASP} code.\cite{VASP,KresseVasp1,KresseVasp2,KresseVasp3}
 using well converged plane wave cut-off energy (500 eV) in the projector augmented wave (PAW)\cite{Blochl94}  method. Supercells of 128 atoms were used to model the defects.

 Unfortunately, the {\sc GIPAW} code has not yet implemented hybrid functionals
 but can since recent improvements include Hubbard-U terms. It is integrated with
 the {\sc Quantum Espresso} (QE) code \cite{QE-2009}  which provides similar functionality to the
 {\sc VASP} code. After determining the self-consistent potential of the system with
 a standard QE run, the {\sc GIPAW} code evaluates the first-order
 induced current using density functional perturbation  (DFPT)
 and from this extracts various $g$-tensor contributions, including
 the magnetic field induced by the first-order current from the Biot-Savart law,
 which leads to the spin-other-orbit contributions.
 It also includes other relativistic corrections besides spin-orbit coupling
 and distinguishes paramagnetic and diamagnetic contributions as explained in detail
 in Ref. \onlinecite{Pickard01,Pickard02}.
 For most of the $g$-tensor calculations we use the generalized gradient approximation
 (GGA) in the Perdew-Burke-Ernzerhof (PBE) parametrization \cite{PBE} but at the
atomic positions relaxed with hybrid functional. 
In some cases we also used Hubbard-U corrections to the PBE functional
which can simulate the hybrid functional effects in creating an orbital
dependent potential with stronger hole localization.  
 We checked that the hybrid functional with the parameters used here, satisfy 
 the generalized  Koopmans' theorem \cite{Mori-Sanchez06,Cococcioni05,Dabo2010,Lany2010} quite well for all of the
 defects considered here and at the same time provide an accurate band gap
 of 3.4 eV. 
 Details about these tests are given
 in Supplemental Material\cite{supinfo}.   We use these calculations to evaluate defect transition
 levels using the standard defect formation energy formalism as outlined in Freysoldt \etal\cite{Freysoldt2014}
 and to determine the
 structural models of the defects. 

 These same relaxed structures
are then used to  calculate $g$-tensors either in PBE or in PBE+U.
 Adding Hubbard-U terms to adjust to hybrid functional results is not trivial.
 One has several choices: adding $U$ on Zn-$d$, O-$p$ and N-$p$. Our aim is
 not to provide a fully optimized choice but to gain insight in qualitative
 effects of adding specific $U$ terms.

 Hyperfine tensors were also calculated using the {\sc GIPAW} code.
 They make use of the PAW reconstruction of the full atomic wave function
 including relativistic corrections.\cite{VdWalleBlochl93,vanLenthe93,vanLenthe94}
 \section{Results}
 \subsection{N$_2$ radicals} \label{sec:radical}
 \subsubsection{$g$-tensors} 
 We start with the results for the $g$-tensors of the N$_2$ molecule in the $+1$ and $-1$ charge states as shown in Table \ref{tabn2+}, \ref{tabn2-}. 
 \begin{table}
   \caption{$\Delta g$-tensor of N$_2^+$ radical in ppm: comparison with other
     calculations and contributions to GIPAW (PBE) as detailed in Ref. \cite{Pickard02}\label{tabn2+}}
   \begin{ruledtabular}
     \begin{tabular}{lrr}
     contribution  & $\Delta g_\parallel$ & $\Delta g_\perp$ \\ \hline
     total  GIPAW (PBE) & -121  & -3180 \\
     total GIPAW (PBE+U) & -125 & -3126 \\
       total Bruna\footnote{Bruna and Grein\cite{Bruna08}}     & -249  & -2734 \\
       total TB model \footnote{Boonchun and Lambrecht\cite{Boonchun13}}  & 0     & -2600 \\ \hline
       relativistic mass    & -259  & -259 \\
       SO bare          &  49 & -704 \\
       SO para          &  0.2  & -2271 \\
       SO dia           &  8     & 12 \\
       SOO              &  81  & 42 \\
     \end{tabular}
   \end{ruledtabular}
 \end{table}
 We can see from the table that the dominant contributions to the $\Delta g_\perp$ are the spin-orbit (SO) paramagnetic and bare term. The bare term refers to the pseudo part of the wave function and the paramagnetic part corresponds to the  PAW reconstructed parts of the full atomic wavefunction. The diamagnetic and spin-other-orbit (SOO) contributions are small.
 The agreement with the other calculations which use a much simpler
 approach is excellent. The Bruna and Grein
 approach \cite{Bruna08}  calculates
 first-order contributions to $\Delta g$ at the realistic open-shell
 Hartree-Fock level and the second order terms correspond to the
 cross terms of orbital Zeeman and spin-orbit coupling, essentially as
 in Eq.(\ref{eqgper}).
 In  the GIPAW approach
the relativistic
 mass term and diamagnetic terms are also first-order terms in the sense that they  are calculated from expectation values
 using the zero-th order wave functions.  The SOO and SO para and bare
 term terms are second order corrections to the energy since they involve
 first-order wave functions. Thus, the sum of SO dia and relativistic mass
 corrections should be compared with the Bruna value for the
 $\Delta g_\parallel$.  The orientation averaged $\Delta g$
 in our present GIPAW calculation is $-2160$ ppm and is close to the experimental value of $-1900$ as reported in Bruna and Grein \cite{Bruna08}.
 The above values were obtained in PBE.  When adding a $U$ on N-$p$ orbitals 
 of 3 eV, the values change slightly. We only report the  decomposition in
 partial contributions for the PBE case. The decomposition is similar
 with the SO diamagnetic mass corrections and SOO almost unchanged and the
 differences arising mostly from the SO paramagnetic and bare terms
 which indeed depend on energy level splittings because they
 are second-order corrections to the energy. This calculation provides
 a good benchmark for the accuracy of the GIPAW approach.

 \begin{table} \caption{Calculated $\Delta g$ (in ppm) terms  for $N_2^-$
     using the GIPAW approach.\label{tabn2-}}
   \begin{ruledtabular}
     \begin{tabular}{lrr}
       contribution & $\Delta g_\parallel$ & $\Delta g_\perp$ \\ \hline
       total GIPAW & 53 & 1741 \\ \hline
       relativistic mass & -2 & -2 \\
       SO bare           & 14 & 447 \\
       SO para           & 41 & 1296 \\
       SO dia            & 0.1  & 0.1  \\
       SOO               & -0.2 & -0.4 \\
     \end{tabular}
   \end{ruledtabular}
 \end{table}
 For the N$_2^-$ radical we find again that the dominant contribution
 are the paramagnetic and bare SO terms. They are positive in this case
 and this is easy to understand from the Eq.(\ref{eqgper}) since now
 the unpaired spin is in the $\pi_g$ state and it can give non-zero off-diagonal matrix elements of the angular momentum operator with
 two lower lying $\sigma_g$ states.  In this calculation, we have
 occupied one of the degenerate $\pi_g$ states and thereby
 broken the symmetry in our spin-polarized DFT calculation. This confirms
 that the N$_2^+$ and N$_2^-$ radicals have opposite signs of the  main
 $\Delta g_\perp$. Of course these results correspond to the isolated
 molecule and this may change when the molecule is placed in a crystal
 environment and other levels of the system become involved. 

 \subsubsection{Hyperfine tensors}
 The hyperfine tensor for the diatomic molecule contains a dipole part which is axial with
 parameters $A_\perp=A_{dip}$, $A_\parallel=-2A_\perp$ and the isotropic Fermi contact term $A_{iso}$.
 Our calculated values compared to experiment and other calculations are given in Table \ref{tabhypn2}.

 \begin{table}
   \caption{Hyperfine tensor parameters for N$_2^+$ and N$_2^-$ in MHz.\label{tabhypn2}}
   \begin{ruledtabular}
     \begin{tabular}{lcc}
       & $A_{dip}$ & $A_{iso}$ \\ \hline
       & \multicolumn{2}{c}{N$_2^+$}\\ \hline
       This work PBE &  $-30.4$       &   102 .2\\
        This work PBE$+U$ & $-29.9$   &   95.6 \\
       Expt     &  23.3          &  102.4\footnote{Scholl \etal \cite{Scholl1998}} \\
                &               &  104.1\footnote{Knight \etal \cite{Knight1983}} \\
       Other calc. & 29.7      & 91.3 \footnote{Bruna and Grein \cite{Bruna08}}\\ \hline
       & \multicolumn{2}{c}{N$_2^-$} \\ \hline
       This work PBE & $-0.14$ & 2.92 \\
       This work PBE$+U$ & $-0.13$ & 2.99\\  
     \end{tabular}
   \end{ruledtabular}
 \end{table}
 Note that the experiment does not detect the sign of the hyperfine tensor.   The agreement is quite good.
 The Fermi contact term depends slightly on the functional. Interestingly, while adding $U$ is expected to make the
 wave function more localized, its Fermi contact term nonetheless slightly decreases. This must indicate that the $s$-component of the  wave function is slightly decreased.  Note that we included $U$ on the N-$2p$. 
 Various other calculated results are reviewed in Bruna and Grein\cite{Bruna08} and give a range of values
 with average 88$\pm10$ for the Fermi contact term.
 For the N$_2^-$ radical, we find a much smaller hyperfine interaction. For the isotropic Fermi contact term, this is clearly
 related to that the unpaired spin in this case is in a $\pi_g$ state and has no direct $s$ contribution to the wave function.

 \subsection{Zn-vacancy} \label{sec:znvac}
 \subsubsection{$g$-tensor}
\begin{figure}
 \includegraphics[width=\linewidth]{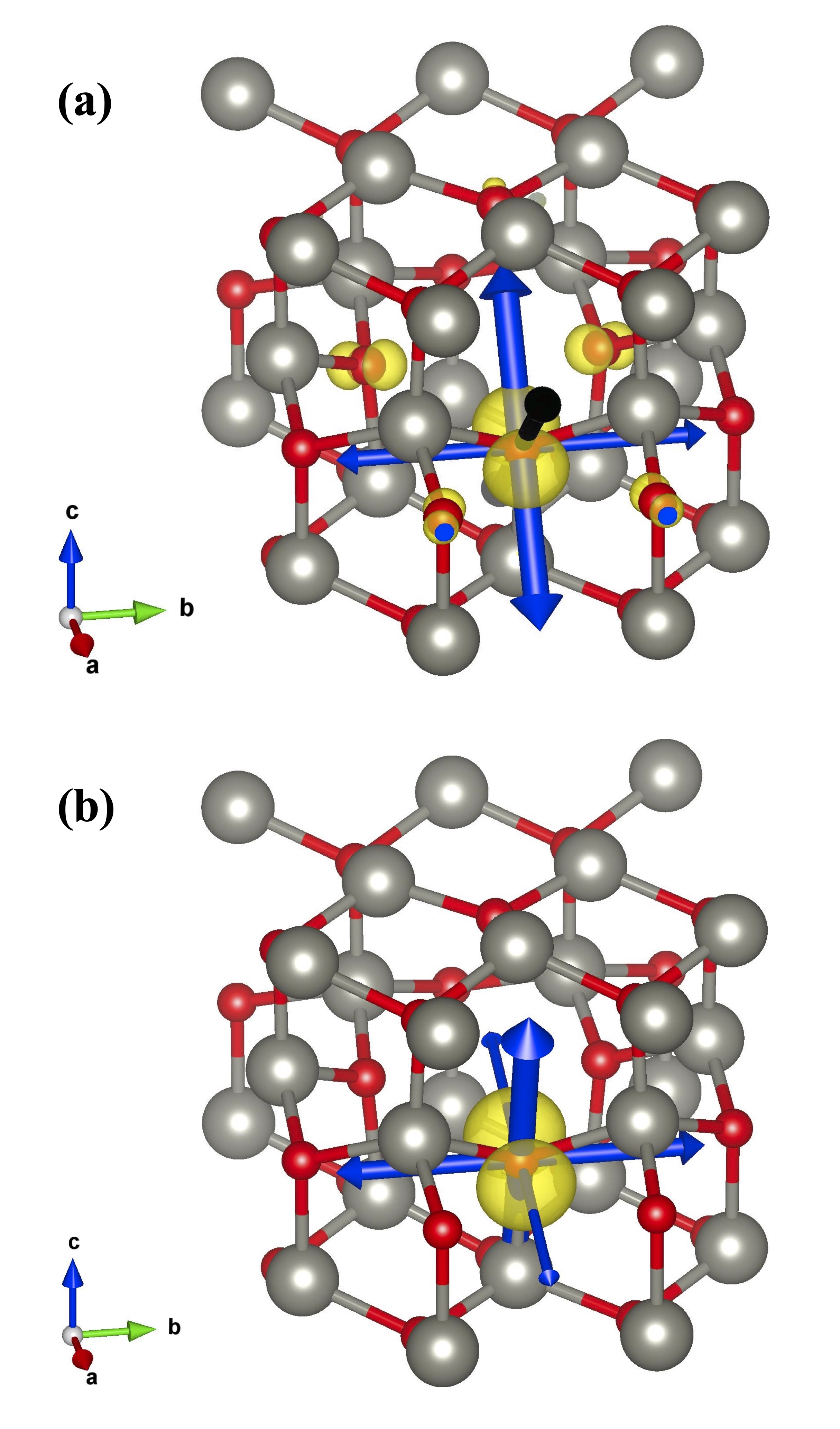}
   \caption{Relaxed structure in (a) PBE and (b) PBE+U
     near the $V_\mathrm{Zn}$ in the EPR active $q=-1$ charge state, showing the
     net spin density as a yellow isosurface. The double-sided vectors show the $\Delta g$ tensor principal axes and the thickness of vectors indicate how big the magnitude of $|\Delta g|$ is, black and blue represents negative and positive values, respectively. \label{figvzn}}
 \end{figure}

 Next we consider the Zn-vacancy.  Unconstrained relaxations carried out in
 hybrid functional led to a model in which the spin is clearly
 localized on a single O which moved away from the vacancy,
 thus forming a polaronic state. This can be seen in Fig. \ref{figvzn}.
  In this case, it was localized on a lateral O in the basal plane next to the vacancy
  and thus the system has only $C_s$ symmetry, containing only  a  mirror plane. The {\bf a}, {\bf b}, {\bf c}
  vectors in this figure are the lattice vectors of the supercell and correspond to $[01\bar{1}0]$, $[\bar{2}110]$
  and $[0001]$ directions. Thus the spin density is seen to lie in a $(\bar{2}110)$ plane and with
  the smallest ($\Delta g<0$) principal axis closer to the {\bf c} axis.   On the other hand, if the spin localizes on the
  axial O, the symmetry of the system remains $C_{3v}$. 
Experimentally both of these cases have been observed in
 the work of Galland and Herve \cite{Galland1970} and the $V_\mathrm{Zn}$
 EPR center was also studied by Son \etal\cite{Son2013}, who identified
 a separate center with H attached to the O in the $V_\mathrm{Zn}$.
 Here, we only discuss the $V_\mathrm{Zn}$.   The $g$-tensor and its principle axes are given in
 Table \ref{tabvzn}.

   \begin{table}
     \caption{$g$-tensor and principle axes for the $V_\mathrm{Zn}$.\label{tabvzn}}
   \begin{ruledtabular}
     \begin{tabular}{lccc}
       & $g_1$ & $g_2$ & $g_3$ \\ \hline 
       Expt.\footnote{Galland and Herve \cite{Galland1970}} & 2.0028 & 2.0173 & 2.0183 \\
       principle axes & $\theta_c=69.25^\circ$   & $\theta_c=20.75^\circ$  &  \\
                      & \multicolumn{2}{c}{in $\langle1\bar{2}10\rangle$} & $\perp \langle 1\bar{2}10\rangle$ \\ \hline
       Calc PBE & 1.9948 & 2.0166 & 2.0096 \\ 
       principle axes    &  $\theta_c=38^\circ$ & $\theta_c=52^\circ$ & \\
       & \multicolumn{2}{c}{in $\langle1\bar{2}10\rangle$} & $\perp \langle 1\bar{2}10\rangle$ \\ \hline
       Calc PBE+U        & 2.0039 & 2.0095 & 2.0092 \\
       principle axes    &  $\theta_c=73^\circ$ & $\theta_c=17^\circ$ & \\
         & \multicolumn{2}{c}{in $\langle1\bar{2}10\rangle$} & $\perp \langle 1\bar{2}10\rangle$ \\
     \end{tabular} 
   \end{ruledtabular}
   \end{table}

   We can see that in the experiment, the smallest $g$ component is still larger than the free-electron value
   $g_e=2.002319$ and has its principle axis at 69.25$^\circ$ from the {\bf c} axis in a $\langle 1\bar{2}10\rangle$
   plane, which is a mirror plane of the wurtzite structure. Note that the opposite direction is 111$^\circ$ from the {\bf c}-axis, which is close to the 109$^\circ$ ideal tetrahedral angle, which means that this direction is the direction
   of the broken Zn-O bond. Thus, as usual the lowest $\Delta g$ occurs is along the direction of the dangling bond.
   The two other principal values are close to each other and are larger and positive. This is also consistent
   with the $C_{3v}$ center with the hole localized on the axial O in which case the expt. values
   are $g_\parallel=2.0024$ and $g_\perp=2.0193$. 
   Comparing to the PBE calculated values, we see the $g$-tensor still lies in a $\langle1\bar{2}10\rangle$ plane but
   the smallest value is now negative and at 38$^\circ$ or 142$^\circ$ from the {\bf c}-axis. Adding a  Hubbard-$U$ term of
   5 eV on the O-$p$ orbitals makes this $\Delta g$ positive, but overshoots slightly compared to experiment.
   The angle $\theta_c$
   from the {\bf c}-axis is now 73$^\circ$, or 107$^\circ$, much closer to the experiment and to the dangling bond
   direction. The wave function also become more localized exclusively on this one O (as shown in Fig. \ref{figvzn}(b), while in PBE it had some small components on the other two lateral O neighbors of the Zn-vacancy. 
   This is as expected from DFT+U in which the $U$ terms
   tends to make the spin density more localized by pushing the hole state deeper into the gap. The principal values
   in the directions perpendicular to the dangling bond are smaller than in experiment but indeed larger than along the
   dangling bond and closer to each other than in the PBE case.
   Using a smaller value of $U=3$ eV give $g_1=2.0054$, $g_2=2.0117$ and $g_3=2.0128$, giving a larger overestimate
   of the small $g_1$, which we might call $g_\parallel$  (meaning parallel to the dangling bond) and larger values for
   $g_2$ and $g_3$ which we might average to $g_\perp$, which are closer to experiment. 
   Further inspection of the $\Delta g$ contributions
   shows that the SOO contribution is small and increasing $U$ increased the paramagnetic SO contribution for
   both parallel and perpendicular directions.   Adding a $U_d=6$ eV on Zn-$d$ and $U_p=3$ eV on O-$p$ reduced
   $g_\parallel$ to 2.0036 but also reduced $g_\perp$ to 2.0097.  The directions of the principal axes
   barely changed.

   These results confirm the basic model proposed by Galland and Herve \cite{Galland1970}, who analyzed the $\Delta g$ tensor
   essentially based on Eq.(\ref{eqgper}) and viewed it as originating from the splitting between the O-$p$
   state on which the hole is localized from its perpendicular directions. Since the hole is a localized
   O-$p$ type dangling bond, it lies above the other O-$p$ states and the SO contribution to $\Delta g$ is thus
   positive for $g_\perp$ and negligible for $g_\parallel$ in their model.  The full calculation indicates that
   the $\Delta g_\parallel$ is not exactly zero and also  slightly positive.  The details depend obviously
   sensitively on the degree of localization of the wave function.  All the models considered here including $U$
   give better results than the pure PBE results because the latter has a wave function too delocalized on
   other nearby O next to the vacancy even though we already created some difference between the three lateral
   O by relaxing the structure within hybrid functional.

   Our HSE and PBE+U calculations show that the defect transition levels for $V_\mathrm{Zn}$ are 1.38 and 0.29 eV, respectively. The HSE results agrees with previous studies of 1.4 eV\cite{Svensson2017,Svensson2018}. The PBE+U gives
   a significantly less deep level, consistent with previous research which
   gives values ranging  0.17--0.3 eV\cite{Shengbai2015,VandeWalle2000,Klein2006}.

   \subsubsection{Hyperfine}
   For the $V_\mathrm{Zn}$ the spin localizes on a single oxygen.
   Oxygen has only isotope $^{17}O$ with non-zero nuclear spin  $I=5/2$
   and this isotope has only 0.038 \%  natural abundance.
   Nonetheless, calculating it gives a  value of $A_{dip}=74$ MHz and
   $A_{iso}=27$ MHz on the oxygen  on which the spin is localized.

   For Zn there is one isotope $^{67}$Zn with spin $I=5/2$ which has 4.10 \%
   abundance. We find  non-negligible Fermi contact terms hyperfine on only
   three of the Zn atoms in the cell, namely the three that are nearest
   neighbors to the oxygen on which the spin is localized. Their hyperfine
   tensors $A_{iso}$ range from -13.7 to -16.7 MHz.  The dipolar parts
   $A_{dip}\approx1$ MHz. 
   
   \subsection{Substitutional N on O-site}\label{sec:no}
\begin{figure}
     \includegraphics[width=\linewidth]{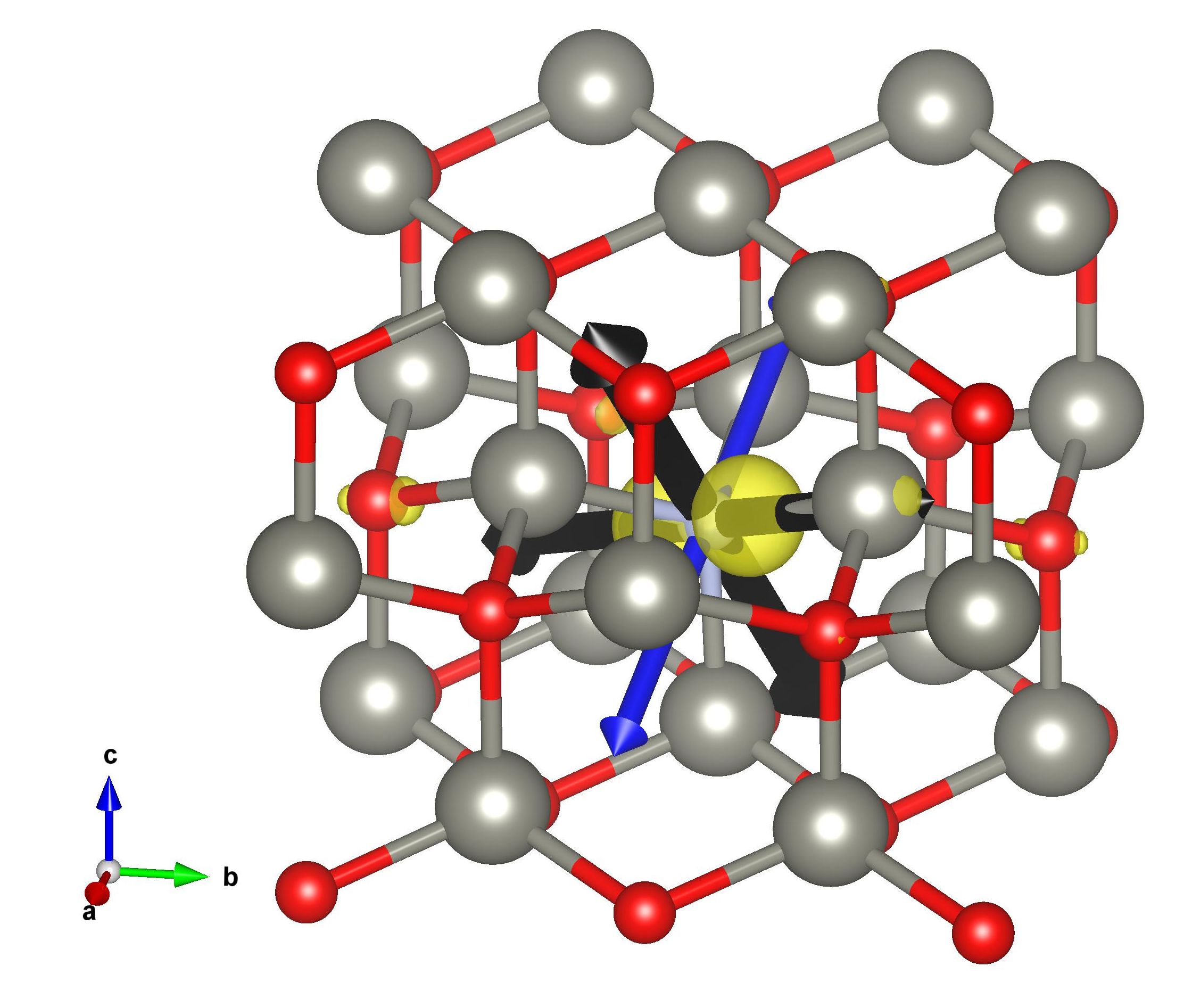}
     \caption{Spin density for N$_\mathrm{O}$ in the neutral charge state and EPR $g$-factor. \label{fign_o}}
   \end{figure}

   Next, we turn our attention to the substitutional N$_\mathrm{O}$ case. This is a well studied defect and
   found to have a very deep $0/-$ level. The results from our HSE calculation of 2.02 eV is a bit deeper than previously obtained values \cite{Boonchun13,lyons2009}. However, our PBE+U functional is 0.56 eV while LDA calculation reported this value of 0.4 eV \cite{Su-Huai2002}.    The spin density of the neutral charge state is shown in Fig. \ref{fign_o}.
   The $g$-tensor is compared with experiment in Table \ref{tabn_o}.
   \begin{table}\caption{$g$-tensors for N$_\mathrm{O}$.\label{tabn_o}}
     \begin{ruledtabular}
       \begin{tabular}{lccc}
         &$g_{\parallel c}$ & $g_{\perp c}$ & \\ \hline
         Expt. \footnote{Phillips \etal \cite{Phillips14}}  & 1.995 & 1.963  & \\ \hline
                       & $g_\parallel$ & $g_c$ & $g_\perp$ \\ \hline 
         PBE                                                & 2.0093 & 1.9810 & 1.9843 \\
         principal axes                                     & $\theta_b=27^\circ$ & $\theta_c=36^\circ$ & $\theta_a=36^\circ$ \\
         PBE+$U_{Np}=3$ eV                                       & 2.0062 & 2.0014 & 2.0029 \\
       \end{tabular}
     \end{ruledtabular}
   \end{table}
   The unconstrained relaxation gave a spin density localized mostly on N on a $p$-orbital approximately along the $b$ direction
   ($[\bar{2}110]$) and with small contributions on various second neighbor O atoms, as can be seen in Fig. \ref{fign_o}.
   In the experiment the $g_\parallel$ corresponds to the {\bf c}-axis and if fully axially symmetric while we find
   a higher anisotropy. At first, one might assume that this just means that the spin became localized on a
   N-$p_z$ orbital along the {\bf c} axis in the experiment.
   However, we may also assume that the experiment sees an average
   of centers with spin localized in the basal plane and along {\bf c}. 
   Also, our $\Delta g_\parallel=6981$ ppm is positive while the experimental value is negative.
   For our calculations $g_\parallel$ indicates parallel to the unpaired spin orbital,  $g_\perp$ indicates
   in the basal plane perpendicular to the spin orbital and $g_c$ indicates along the {\bf c} axis.
   Averaging the values in the {\bf c} direction, assuming three equivalent in the basal plane orientations and
   one along the {\bf c} axis, we can write $\bar{g}_c= (g_\parallel+3g_c)/4$ which gives a value of 1.988.
   For the direction perpendicular to {\bf c} we can write
   $\bar{g}_{\perp c}= [(g_\parallel+ g_\perp)3/2+(g_c+g_\perp)/2]/4=1.993$. This gives two negative $\Delta g$ values
   close to each other as in the experiment but fails to capture the small difference between in the basal plane
   and along {\bf c} observed in the experiment. There might be some energetic advantage to the
   spin localizing in the {\bf c} direction which would then explain the smaller negative value
   in the {\bf c} direction.
   Adding $U$ values of  3 eV or 5 eV on N or both on N and O did not change the orientation of the spin density
   nor its degree of localization. 
   It tends to make the $\Delta g$ values closer to each other and  smaller  but no improvement with the experimental
   values was obtained.

   It was found experimentally\cite{Phillips14} that the N$_\mathrm{O}$
   center is activated by light of about 1.9 eV.
   We have thus calculated the vertical transitions energy from the
   negative charge state to the neutral one plus an electron at the conduction
   band minimum.  Including only the image charge correction to the negative
   charge state, we obtain 1.98 eV for this activation. However, recently,
   it was proposed\cite{Falletta20} that even the neutral charge state
   in the frozen geometry of the negative charge state requires a
   correction due to the presence of polarization charge and this needs
   be screened using only the electronic screening.  This gives 2.25 eV.
   Both are in reasonable agreement with the experiment.

   As for the hyperfine tensor for N$_\mathrm{O}$, we find $A_{iso}=24$ MHz
   and $A_{dip}=-27$ MHz on the N atom
   with the $A_\parallel=-2A_{dip}$ along the {\bf c} axis.
   Thus, we obtain $A_{\parallel c}\approx 78$ and $|A_{\perp c}|\approx3$ MHz.
   Phillips \etal\cite{Phillips14} give values of $A_{\parallel c}=81$ and
   $A_{\perp c}=8.5$ MHz. These values are in fair agreement.

 \subsection{N$_2$ in ZnO} \label{sec:n2zno}
 \subsubsection{Zn-site}\label{sec:n2zn}
 First we considered various models for N$_2$ placed inside the Zn-vacancy. One of our goals here is to revisit
 the question of whether the N$_2$ in this site is a shallow or deep acceptor.  We start from different initial
 orientations of the molecule, either parallel to {\bf c} or in the basal plane and in the\ basal plane either
 with the molecular axis pointing toward one of the neighboring  O or perpendicular to it.  We also started either from
 the ideal crystal or from the previously relaxed vacancy. To summarize these results, we found that the lowest energy
 for the neutral charge state has the N$_2$ forming a bridge like bond along one of the tetrahedral sides surrounding
 the vacancy and connecting to two O atoms. This configuration, also reported by Petretto and Bruneval\cite{Petretto14},
 has about 0.465 eV lower than the in-basal plane configuration with the molecule aligned with one of the
 bonds in which case it can still make a single N-O bond if we place it close to an O or can be essentially isolated.
 The vertically aligned molecule tended to flip back
 to a horizontal position or at least tilt slightly. We also shifted the center of gravity of this molecule
 up or down from the $V_\mathrm{Zn}$ center to keep it more isolated.     On the other hand in the EPR active
 $q=-1$ charge state the isolated N$_2$ molecule had the lowest energy.  The transition level $(0/-)$ is in principle calculated from the lowest energy configuration of each charge state and is then found to be 2.46 eV using the HSE functional which is considerably deeper above the valence band maximum than the  shallow one of 0.17 eV obtained using PBE+U. These results support the conclusion of Petretto and Bruneval\cite{Petretto14} and contradict those of Boonchun
 and Lambrecht\cite{Boonchun13}, which did only consider a more isolated configuration of N$_2$ and furthermore  proposed that
 the generalized Koopmans' condition is better satisfied within PBE than  PBE+U and thereby obtained a shallower defect level.  Thus, our first conclusion here is that N$_2$ on Zn site would be a deep acceptor.
The figure of these structural models can be found in Supplemental Material\cite{supinfo}.

\begin{figure}
   \includegraphics[width=\linewidth]{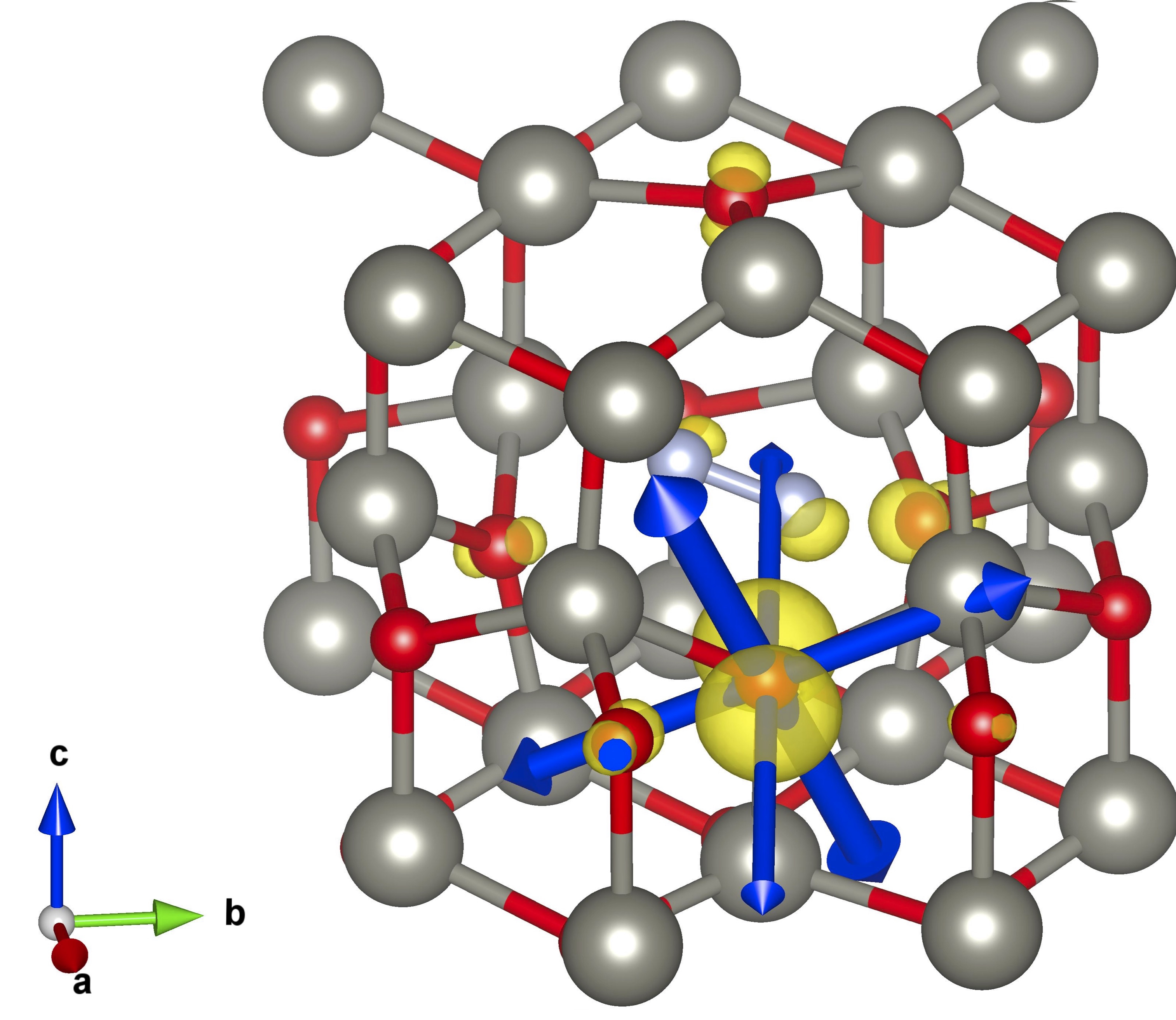}
   \caption{N$_2$ in $V_\mathrm{Zn}$ relaxed structure in the $q=-1$ state with spin isosurface and $g$-tensor.\label{fign2zn}}
 \end{figure}

 Turning now to the spin densities, we find that the spin density showed very little contribution on the N$_2$
 mostly on a single O neighbor. This is shown in Fig. \ref{fign2zn}.
 The $g$-tensor is then similar to the $V_\mathrm{Zn}$ case with all
 positive $\Delta g$ values. Again, if the spin was oriented for example along the {\bf a} direction, then the
 lowest $\Delta g$ principal value occurred along that direction.  The $\Delta g$ shifts are similar to that
 of the vacancy. These are very different from the values reported for N$_2$ in ZnO by Garces \etal \cite{Garces03} and
 Phillips \etal \cite{Phillips14} and we thus conclude that N$_2$ on a Zn site is incompatible with the observed
 EPR center. Furthermore it implies that if N$_2$ would be isolated in a Zn-vacancy it only slightly perturbs
 the vacancy and leads to a polaronic systems with spin localized on a single O as in the vacancy case.
 \subsubsection{Interstitial sites} \label{sec:n2int}
 Next we consider various interstitial sites. Among these, the lowest energy is obtained for the structure where
 N$_2$ occurs in  the middle of the large hexagonal interstitial site.  We constrained this model 
 so as to allow the molecule only to move along the $z$ direction. It is shown in the Supplemental Material \cite{supinfo}. In this case, we found that the N$_2$ molecular states are deep enough that the HOMO $\sigma_{g+}$ state stays occupied and the $\pi_g$ is empty. The neutral charge state shows no levels in the gap at all. Attempting to make a $q=+1$ charge state then leads to removing an electron from the VBM resulting in
 a very delocalized spin density. The $g$-tensor calculated shows a very large negative values along the {\bf c}
 direction, about $g_c=1.7132$ and a value of about $g_{\perp c}\approx1.99$. We think these may reflect the $g$-tensor
 of the VBM but additional work is needed to understand these values.  The $g$-tensor of such delocalized states
 like the VBM or CBM are usually discussed in terms of ${\bf k}\cdot{\bf p}$ theory.\cite{Lambrecht2002}
 We do not discuss it further here but rule out any of these sites as responsible for the observed N$_2$ EPR center
 in ZnO.  Although this does not refute that N$_2$ could occur interstitially as claimed by Nickel and Gluba\cite{Nickel09},
 it presumably has no EPR active state in this case because no defect levels are found in the gap from which
 a singly occupied unpaired spin state can be constructed. Other interstitial forms of N$_2$ may disrupt the
 ZnO network and hence lead to O dangling bond type states as reported by Nickel and Gluba\cite{Nickel09} but they do not lead to an EPR center
 with spin density on the N$_2$ molecule compatible with the one observed \cite{Garces03} and are thus not further pursued here. 
 
 \subsubsection{O-site} \label{sec:n2o}
\begin{figure}
   \includegraphics[width=\linewidth]{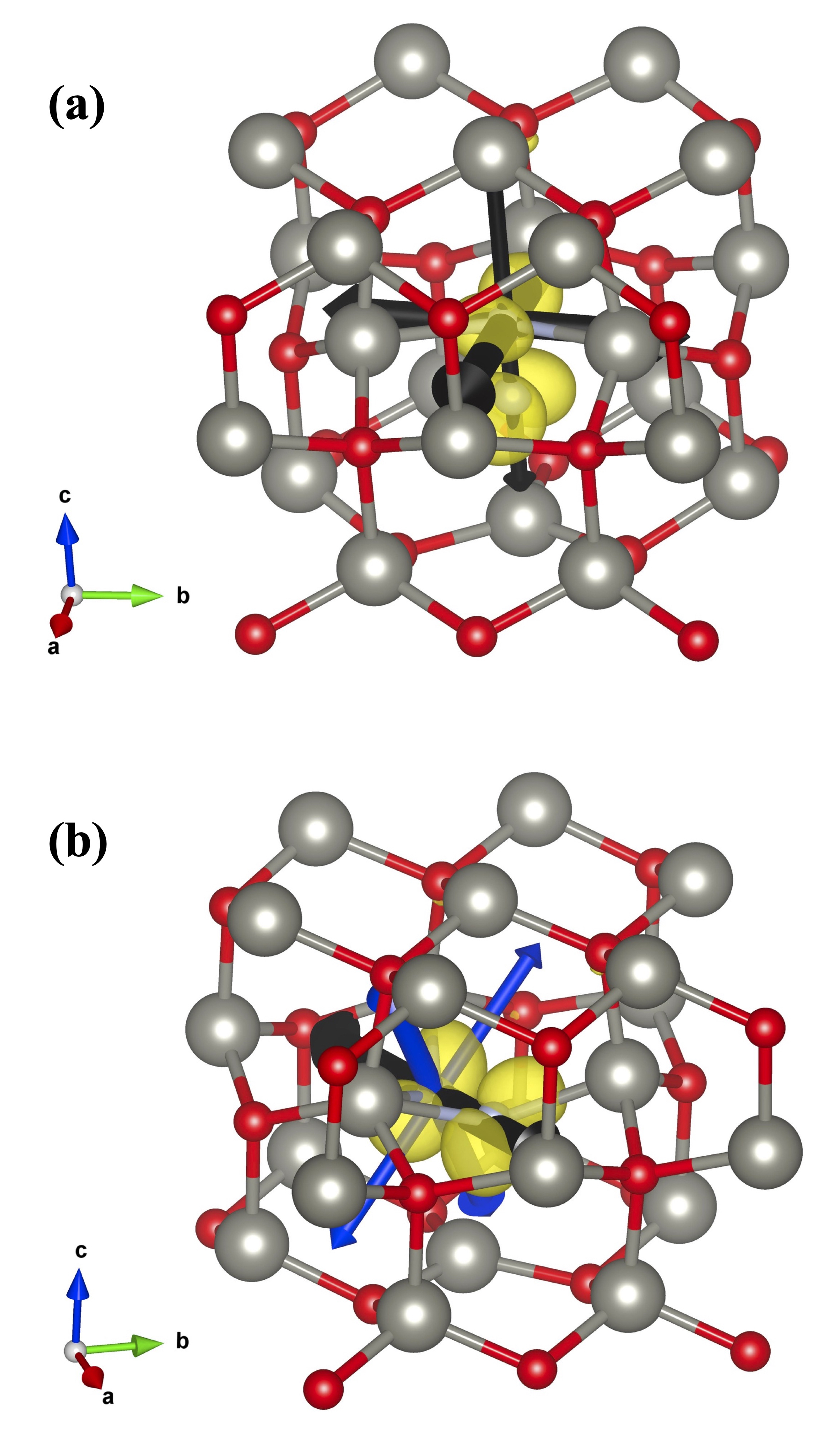}
   \caption{Spin density and relaxed structure for N$_2$ in the O-site with the molecular axis fixed
     (a) parallel and (b) perpendicular to the {\bf c}-axis.
     \label{fign2o}} 
 \end{figure}

 Finally, we return to the N$_2$ molecule in the O-site as initially proposed by Garces \etal\cite{Garces03}.
 We started out from an initial orientation of the molecular N$_2$ bond axis parallel to the {\bf c} axis. The N$_2$ molecule was allowed to move only in {\bf c} direction. After relaxation
 the spin density was strongly localized on the N$_2$ molecule and shows clearly a $\pi_g$ like state which happened to be oriented with the {\bf a} axis. This is shown in Fig. \ref{fign2o}(a). The defect in this case is a donor
 and the spin density corresponds to the $q=+1$ charge state of the defect, which is, however, a N$_2^-$
 from the view of the N$_2$ molecule.
 The defect transition level (+/0) for this case is 3.07 using HSE (1.46 eV  in  PBE+U). 
 We also investigate the  N$_2$ molecular with its bond axis perpendicular to the {\bf c} axis.
 We start the configuration by pointing the N$_2$ toward one of neighboring Zn as can be seen in Supplemental Material\cite{supinfo}. After the relaxation by fixing the molecule's movement in the $z$ direction, the N$_2$ molecule is pointing to the space between two Zn atoms. Like the previous model, the spin density was strongly localized on the N$_2$. The defect transition levels of this model, 3.23 and 1.59 eV using HSE and PBE+U, are somewhat deeper than those of N$_2$ parallel to the {\bf c} axis.  Full unconstrained relaxation  of the N$_2$ molecule  led to an orientation
 intermediate between these two cases.

 \begin{table}\caption{$g$-tensors for N$_2$ on O-site\label{tabn2_o}}
   \begin{ruledtabular}
     \begin{tabular}{llccc}
       && $g_{\parallel \mathrm{N}_2}$ & $g_{\parallel \pi-orbital}$ & $g_{\perp \pi-orbital}$ \\ \hline
     N$_2$  $\parallel {\bf c}$ &PBE  & 1.95378      & 1.98724        & 2.00007 \\ 
       & PBE$+U$\footnote{$U=3$ eV on N and O}           & 1.98736      & 2.00381        & 2.00988  \\
     N$_2$  $\perp {\bf c}$ & PBE     & 1.98494      & 2.00506        & 2.00911  \\
     & PBE$+U$          & 1.99281      & 2.00457        & 2.00666 \\ \hline 
     &         & $g_{\parallel c}$ & $g_{\perp c}$ & \\ \hline
     & Expt.                                 &  2.0036 & 1.9935 \\
     N$_2$ $\perp {\bf c}$ & SOMO $\parallel {\bf c}$   &  2.0038 & 1.9986\footnote{average of columns 1 and 3 for row 2}  \\   
     \end{tabular}
     \end{ruledtabular}
 \end{table}

 The $g$-tensor for the N$_2$ molecule parallel to the ${\bf c}$- axis,  is found to have the lowest principal value along the direction of the N$_2$ axis
 and in fact has a negative $\Delta g$ in this direction. 
 The next higher $g$-principal value is in the direction of the $\pi_g$ SOMO orbital
 and the highest value perpendicular to the plane
 of the molecular axis and the $\pi$-orbital. This corresponds to the first row in Table \ref{tabn2_o}. 
 We can see in rows 3-4 of the table that this is true also when the N$_2$ axis is fixed to be
 perpendicular to {\bf c}. In case of a full relaxation, we find some intermediate orientation. We do not find  convincingly lower
 energy among any of these orientations, which in the end result in energies within the error bar from each other. 
 
 All values slightly increase if we add a Hubbard-$U$ on
 both N and O and this makes the two higher values correspond to a small positive $\Delta g$ for the first case
 were N$_2$ is parallel to the ${\bf c}$ axis.  When we fix N$_2$ to lie in the {\bf c} plane, we already
 obtain two positive and one negative $\Delta g$ in PBE but the SOMO  is not lying in the plane but
 is slightly tilted as shown in Fig. \ref{fign2o}(b).
 There are thus small variations in these $g$-tensors
 with the functional and depending on the orientation of the molecule but the basic correlation between molecular axes,
 and the broken symmetry of the $\pi_g$ orbital in which the hole resides stay consistent. 
 Further inspection shows that,  as usual, the mass correction, diamagnetic and SOO contributions are small.
 The main paramagnetic contribution is strongly negative both in the GIPAW and bare terms for the {\bf c} direction
 while the paramagnetic GIPAW term is positive for the directions in the plane of the molecular axis and its spin
 orbital. 
 
 The $g$-tensor at first does not seem to agree with the experiment which has
 a $g_{\parallel c}=2.0036$ and $g_{\perp c}=1.9935$.
 However, let's now consider that the molecule might be oriented in various equivalent ways and that the
 experiment sees an unresolved average of these.  We then have several possibilities, the molecular axis might be along {\bf c}
 as in row 1 of Table \ref{tabn2_o} or perpendicular to it with either the $\pi$ orbital along {\bf c},
 or perpendicular to it, or somewhere in between as in the last two rows of the table. 
 We obtain the following averages based on the $g$-tensors of row 2 of Table \ref{tabn2_o}. 
 For N$_2$ axis parallel to {\bf c}: $g_{\parallel c}=1.98736$, $g_{\perp c}=2.0068$,
 for N$_2$ axis in the basal plane and the $\pi$ orbital along {\bf c}: $g_{\parallel c}=2.00381$, $g_{\perp c}=1.9986$,
 and finally for N$_2$ axis in the basal plane and the $\pi$ orbital perpendicular  {\bf c}:
 $g_{\parallel c}=2.00988$, $g_{\perp c}=1.99558$.
 The first choice disagrees with experiment but both cases with the axis in the basal plane are compatible
 with the experimental value with a slightly better agreement if the $\pi$-orbital is along the {\bf c} axis.
 In fact, in this case the agreement is pretty close.  Furthermore, also in the two cases (row 3 and 4)  where we
 explicitly constrained the molecular axis to be in the plane but found the SOMO to be tilted away from the plane,
 we find a negative $\Delta g$ perpendicular to the plane along the molecular axis and the other two directions have positive
 $\Delta g$ with the largest one in the direction perpendicular to the plane of the molecular axis and the SOMO orbital.
 This direction is found to be closest to the {\bf c}-direction, about 30 $^\circ$ away from it.
 We may deduced from this that the experimental data are compatible with a preferred orientation of the molecular axis
 in or close to the basal plane. The axial symmetry along {\bf c} observed experimentally does not correspond to a simple orientation of
 the molecule along this axis but rather some average over various in-plane orientations of the molecule.

 \begin{table}
   \caption{Hyperfine parameters (in MHz) on N for N$_2$ on O-site. \label{hypn2_o}}
   \begin{ruledtabular}
     \begin{tabular}{llcc}
       &atom&  $A_{dip}$ & $A_{iso}$ \\ \hline
       N$_2$ $\parallel {\bf c}$ PBE & N$_1$ & $(-28.5,-30.7,59.2)$ & 15.7 \\
                                    & N$_2$ & $(-17.0,-19.9,36.9)$ & 6.3  \\
       N$_2$ $\perp {\bf c}$ PBE    & N$_1$ & $(-23.1, -23.6,46.7)$ & 14.6 \\
                                    &N$_2$  & $(-23.4,-23.9, 47.3)$ & 14.9 
     \end{tabular}
   \end{ruledtabular}
 \end{table}

 The activation energy of the EPR center of (N$_2$)$_\mathrm{O}$ corresponds to a transition from the neutral
 to the positive state releasing an electron to the conduction band minimum. This vertical transition is calculated
 to be 1.57 eV for the case of the molecular axis being in plane and 1.72 eV for the vertically aligned molecular axis,
 following the approach of Falletta \cite{Falletta20} for the correction terms, in other words, using here only
 electronic screening.  These values are in reasonable agreement with the experimental observation that the EPR
 signal of the N$_2$ in ZnO is enhanced by light of already 1.4 eV, so somewhat lower by about 0.5 eV than the
 N$_\mathrm{O}$ substitutional defect.  Since we here found the N$_\mathrm{O}$ calculated to have an activation
 energy of around 2.2 eV, this is qualitatively also in agreement with experiment. 

 Finally, we consider the hyperfine tensor for the N$_2$ on O site.  We give the eigenvalues of the hyperfine dipolar tensor
 as three values in parentheses  for  each of the N atoms.   For the N$_2$  $\parallel{\bf c}$ case, the values on both
 atoms differ somewhat but their average value is still close to $-23$ MHz, which is also close to that of  the isolated
 N$_2^+$  molecule.  For the isotropic part, the value is about 10 times smaller than for the N$_2^+$ case.
 This could at first sight indicate significant delocalization of the defect wave function. However, what really matters for the
 Fermi contact term is the amount of N-$s$ wave function.  
 In fact, the values are significantly larger than those of the N$_2^-$ isolated radical.
 The wave function here is clearly not purely $p$-like on N
 even though it is related to a $\pi_g$ state. 
 The N-$p$-like part of the wave function is responsible for the
 dipolar part and the closeness to the values for the N$_2^+$ molecule indicate that the wave function is still strongly
 localized on the N$_2$. 

 Comparing to experimental data by Garces \etal\cite{Garces03} and Phillips \etal \cite{Phillips14}, who give $A_{\parallel c}=9.8$ and $A_{\perp c}=20.1$ MHz,
 we note that our value for $A_{\parallel N_2}=-8$ and $A_{\perp N_2}=62$ MHz.
 So, if the molecule lies in the plane $|A_{\parallel c}|=8$ MHz 
 and the $|A_{\perp c}|$ is the average of these two or 27 MHz. These are
 consistent with the experimental values.

\subsubsection{Summary}
 We thus conclude that among the various models for N$_2$ in ZnO, only the O site gives possibly
 a $g$-tensor compatible with the experiment because it is the only model with spin density
 localized on the N$_2$ molecule. In order to obtain agreement we need to assume that the N$_2$ molecule tends to lie
 preferentially with
 its axis in the basal plane and most likely with the $\pi$-orbital  containing the unpaired spin
 pointing in the {\bf c} direction or close to it.
 In fact, there are three possible high-symmetry orientations in the plane  with the molecular axis in a mirror plane
 vs. only one along {\bf c}  for the N$_2$ axis, so simply statistically, it is more likely to find planar orientation.  We did
 not find a clear energy advantage for this orientation. They were all close and also close to the fully relaxed minimum
 energy orientation which was intermediate. Thus we assume that the experiment samples some average over these different
 orientation of the molecule. The degeneracy of the $\pi_g$ state is broken with one particular orientation of the orbitals
 containing the unpaired spin. 
 Because the $g$-tensor in the direction of the bond is strongly negative, the dominant in-plane orientation of the molecular axis  leads to an average negative 
 $\Delta g_{\perp c}$  value in agreement with experiment.  The otherwise mostly positive $\Delta g$ values are
 compatible with our initial calculation for the N$_2^-$ radical and the largest positive value
 occurs for the direction perpendicular to the plane of the SOMO, which we found to be close the c-axis, thus
 explaining the positive $\Delta g_{\parallel c}$ in the experiment. 
 The negative value along the bond
 must arise somehow from the interplay with the crystal levels rather than from the molecule itself.

 \section{Conclusions}

 In conclusion, we have carried out $g$-tensor calculations for isolated molecules of N$_2$ radicals with
 one electron subtracted or added and analyzed the different contributions to it in the GIPAW theory
 and compared them to previous perturbation theory approaches.  We have shown that the EPR signal of
 N$_2$ in ZnO  is only compatible with calculated $g$ tensors for the O-site. In that case, the N$_2$ behaves
 as a deep donor and this is compatible with the recharging studies of Phillips \etal. \cite{Phillips14}.
 For the Zn-site, the N$_2$ molecule tends to bind to two O in the neutral state but stays in an isolated non-bonding
 configuration in the unpaired spin negative state. The system is then unfortunately a deep acceptor.
 This agrees with Petretto and Bruneval's study \cite{Petretto14}. The spin density in this case is rather similar to that of the Zn-vacancy for which we found good agreement for the $g$-tensor with early experimental data by
 Galland and Herve \cite{Galland1970} characterized by a positive $\Delta g_\perp$-tensor where perpendicular means perpendicular to the dangling bond. 
  For the interstitial sites, no levels in the gap
  are obtained and hence no spin density is observed unless we remove an electron from the VBM which gives a
  very different $g$-tensor.    For the simple substitutional N$_\mathrm{O}$  we also found a $g$-tensor
  in reasonable agreement with experiment assuming that the experiment sees an average over different possible orientations of the N-$p$ orbital on which the spin is localized.
  These results suggest that further experimental work on these EPR centers, possibly with higher microwave frequency
  and magnetic field could help to resolve the individual centers with different orientation of the spin density.

  Good agreement with experiment is also obtained for the hyperfine parameters
  both in the isolated molecules and for the N$_2$ molecule on the O
  site and for the N$_\mathrm{O}$ substitutional case. We also provided
  hyperfine parameters for the $V_\mathrm{Zn}$ where we find notable
  hyperfine parameters only on the three nearest neighbor Zn to the O
  on which the electron spin is localized.\\

  \acknowledgements{We thank Davide Ceresoli and Dmitri Skachkov for useful discussions. K.D. was supported by the National Research Council of Thailand (NRCT), Grant No. NRCT5-RGJ63002-028. A.B. has been funded by National Research Council of Thailand (NRCT; 153/2564).  W. R. L. L was supported by the U.S. Department of Energy Basic Energy Sciences (DOE-BES) under Grant No. DE-SC0008933.  The calculations were performed at the Ohio Supercomputer Center.}
  
 \bibliography{gZnO}
\end{document}